\documentclass[11pt]{article}
\usepackage{amssymb,amsmath,latexsym}
\usepackage{mathrsfs}
\usepackage{tipa}
\usepackage{amsfonts}
\usepackage{graphicx}
\usepackage{subfigure}
\usepackage[title]{appendix}
\usepackage{color}
\usepackage{multirow}
\usepackage{amsbsy}
\usepackage{longtable}
\usepackage{floatrow}
\usepackage{indentfirst}
\usepackage{cite}
\usepackage{paralist}
\allowdisplaybreaks[4] 
\newtheorem{theorem}{Theorem}
\newtheorem{definition}{Definition}
\newtheorem{cor}{Corollary}
\usepackage[font=small]{caption}

\begin{document}
	\title{\bf Recovery analysis for weighted mixed $\ell_2/\ell_p$ minimization with $0<p\leq 1$}
	\author {Zhiyong Zhou\footnote{Corresponding author, zhiyong.zhou@umu.se.}, Jun Yu\\
		Department of Mathematics and Mathematical Statistics, Ume{\aa} University, \\Ume{\aa},
		901 87, Sweden}
	\maketitle
	\date{}
	\noindent
$\mathbf{Abstract}$: We study the recovery conditions of weighted mixed $\ell_2/\ell_p\,(0<p\leq 1)$ minimization for block sparse signal reconstruction from compressed measurements when partial block support information is available. We show that the block $p$-restricted isometry property (RIP) can ensure the robust recovery. Moreover, we present the sufficient and necessary condition for the recovery by using weighted block $p$-null space property. The relationship between the block $p$-RIP and the weighted block $p$-null space property has been established. Finally, we illustrate our results with a series of numerical experiments. \\
$\mathbf{Keywords}$: Compressive sensing; Prior support information; Block sparse; Non-convex minimization.

\section{Introduction}
Since its advent \cite{crt,ct1,ct2,d}, Compressive Sensing (CS) has attracted considerable attentions (see the monographs \cite{ek,fr} for a comprehensive view). CS aims to recover an unknown signal with underdetermined linear measurements. Specifically, the standard CS problem consists in the reconstruction of a $N$-dimensional sparse or compressible signal $x$ from a significantly fewer number of linear measurements $y=Ax+e\in\mathbb{R}^m$ with $m\ll N $ and the noise $e$ satisfying $\lVert e\rVert_2\leq \varepsilon$ for some known constant $\varepsilon>0$. In fact, if the measurement matrix $A$ satisfies the restricted isometry property (RIP) condition, the robust recovery can be guaranteed by using the $\ell_1$ minimization (Chapter 6, \cite{fr}) \begin{align}
\min\limits_{z}\,\,\lVert z\rVert_1,\,\,\,\text{subject to $\lVert y-Az\rVert_2\leq \varepsilon$}.
\end{align}

To obtain better recovery performance, the structures and prior information of signals are incorporated in the recovery algorithms. In this paper, we consider both cases. \\ 
\indent
As for the structure, we assume that the unknown signal $x$ is block sparse or nearly block sparse \cite{de,ekb,em,er,ev}, which means that the nonzero entries of $x$ occur in clusters. Block sparse model appears in many practical scenarios, such as when dealing with multi-band signals \cite{me2}, in measurements of gene expression levels \cite{pvmh}, and in colour imaging \cite{mw}. Moreover, block sparse model can be used to treat the problems of multiple measurement vector (MMV) \cite{ch,em,me1} and sampling signals that lie in a union of subspaces \cite{bd2,em,me2}. With $N=\sum_{i=1}^{n}d_i$, we define the $i$-th block $x[i]$ of a length-$N$ vector $x$ over $\mathcal{I}=\{d_1,\cdots,d_n\}$. The $i$-th block is of length $d_i$, and the blocks are formed sequentially so that
\begin{align}
x=(\underbrace{x_1\cdots x_{d_1}}_{x^{T}[1]}\underbrace{x_{d_1+1}\cdots x_{d_1+d_2}}_{x^{T}[2]}\cdots\underbrace{x_{N-d_n+1}\cdots x_N}_{x^{T}[n]})^T. \label{signal}
\end{align}
Without loss of generality, we assume that $d_1=d_2=\cdots=d_n=d$, implying that $N=nd$. A vector $x\in\mathbb{R}^N$ is called block $k$-sparse over $\mathcal{I}=\{d,\cdots,d\}$ if $x[i]$ is nonzero (i.e., $\lVert x[i]\rVert_2>0$) for at most $k$ indices $i$. In other words, by denoting
$\lVert x\rVert_{2,0}=\sum_{i=1}^{n}I(\lVert x[i]\rVert_2>0)$,
a block $k$-sparse vector $x$ can be defined by $\lVert x\rVert_{2,0}\leq k$. For any $p>0$, we define the mixed $\ell_2/\ell_p$ norm $\lVert x\rVert_{2,p}=(\sum_{i=1}^n\lVert x[i]\rVert_2^p)^{1/p}$. To make explicit use of the block structure to achieve better sparse recovery performance, the corresponding extended version of sparse representation algorithm has been developed, namely the mixed $\ell_2/\ell_1$ minimization, \begin{align}
\min\limits_{z}\,\,\lVert z\rVert_{2,1},\,\,\,\text{subject to $\lVert y-Az\rVert_2\leq \varepsilon$}.
\end{align}
It was shown in \cite{em} that if the measurement matrix $A$ satisfies the block RIP condition which generalizes the
conventional RIP notion, then the mixed $\ell_2/\ell_1$-norm recovery algorithm is guaranteed to recover any block sparse signal, irrespectively of the locations of the nonzero blocks. Furthermore, recovery will be robust in the presence of noise and modeling errors (i.e., when the vector is not exactly block sparse). Other existing recovery algorithms include group lasso \cite{yl}, adaptive lasso \cite{lbw}, iterative reweighted $\ell_2/\ell_1$ recovery algorithms \cite{zb}, block version of the CoSaMP algorithm \cite{ekb}, and the extensions of the CoSaMP algorithm and of the Iterative Hard Thresholding algorithm \cite{bcdh}.

On the other hand, we also consider the case that an estimate of the block support of the signal is available. The related literatures on signal recovery using partial support or block support information include \cite{fmsy,hwwx,j,kxah,nsw,inw,rw,sac,vl,yb}. For an arbitrary signal $x\in\mathbb{R}^N$ defined as (\ref{signal}), let $x^k$ be its best approximation with $k$ nonzero blocks, so that $x^k$ minimizes $\lVert x-g\rVert_{2,1}$ over all block $k$-sparse vectors $g$. Let $T_0$ be the block support of $x^k$, where $T_0\subset\{1,\cdots,n\}$ and $|T_0|\leq k$. Let $\tilde{T}$, the block support estimate,  be a subset of $\{1,2\cdots,n\}$ with cardinality $|\tilde{T}|=\rho k$, where $0\leq \rho\leq a$ for some $a>1$ and $|\tilde{T}\cap T_0|=\alpha\rho k$ (for interpretation of $\rho$ and $\alpha$ see Remark 1 in Section 2). To incorporate prior block support information $\tilde{T}$, the weighted mixed $\ell_2/\ell_1$ minimization \begin{align}
\min\limits_{z}\sum\limits_{i=1}^n w_i\lVert z[i]\rVert_2,\,\,\,\text{subject to $\lVert y-Az\rVert_2\leq\varepsilon$},\,\,\,
\text{where $w_i=\begin{cases}
	\omega \in[0,1], &\text{$i \in\tilde{T}$}\\
	1, &\text{$i\in\tilde{T}^c$}
	\end{cases}$}
\end{align}
is adopted. The main idea is to choose $\omega$ such that the $\ell_2$ norm of the blocks of $x$ that are "expected" to be large are penalized less in the weighted objective function.

Moreover, as is shown in many literatures \cite{c,fl,xcxz}, $\ell_p$-minimization with $0<p<1$ allows exact recovery with fewer measurements than that by $\ell_1$-minimization. Thus, it is natural to adopt the nonconvex minimization to the setting of block sparse signal reconstruction with prior block support information. Specifically, we consider the weighted mixed $\ell_2/\ell_p\, (0<p\leq 1)$ minimization problem:
\begin{align}
\min\limits_{z}\sum\limits_{i=1}^n w_i\lVert z[i]\rVert_2^p,\,\,\,\text{subject to $\lVert y-Az\rVert_2\leq\varepsilon$},\,\,\,
\text{where $w_i=\begin{cases}
	\omega \in[0,1], &\text{$i \in\tilde{T}$}\\
	1, &\text{$i\in\tilde{T}^c$}
	\end{cases}$}. \label{min}
\end{align}

When there is no prior block support information ($\omega=1$), the mixed $\ell_2/\ell_p (0<p\leq 1)$ minimization problem has been studied in \cite{wwx1,wwx2}. And the case that there is partially known signal block support but with $\omega=0$ was considered in \cite{hwwx}. We generalize the existing results to incorporating the prior known block support information with a weight $\omega\in[0,1]$. In summary, the main contributions of our work include the following aspects. First, we provide the recovery analysis for the weighted mixed $\ell_2/\ell_p\,(0<p\leq 1)$ minimization by using block $p$-RIP condition. This result extends the existing literatures \cite{fmsy,hwwx,vl,wwx1,wwx2}. Second, we propose the weighted block $p$-null space property and present the sufficient and necessary condition for the weighted mixed $\ell_2/\ell_p\,(0<p\leq 1)$ minimization by this new tool.
Third, we establish the relationship between block $p$-RIP condition and weighted block $p$-null space property. Finally, we illustrate our results via a series of simulations.

The paper is organized as follows. In Section 2, we present the sufficient recovery condition by using block $p$-RIP. In Section 3, we introduce the notion of weighted block $p$-null space property (NSP) and establish the sufficient and necessary condition with this new tool. In Section 4, we establish the relationship between these two conditions. In Section 5, we conduct some simulations to illustrate the theoretical results. Section 6 is devoted to the conclusion.

\section{Block $p$-RIP}
As for the weighted mixed $\ell_2/\ell_p\,(0<p\leq 1)$ minimization, we can obtain the reconstruction guarantees by using block $p$-RIP. We begin with introducing the definition of block restricted $p$-isometry constant, which is a natural extension of the conventional restricted $p$-isometry constant.\\

\begin{definition}[\cite{cs,hwwx,wwx1,wwx2}] Given a measurement matrix $A\in\mathbb{R}^{m\times N}$ and $0<p\leq 1$. Then the block $p$-restricted isometry constant (RIC) $\delta_{k|\mathcal{I}}$ over
$\mathcal{I}=\{d_1,\cdots,d_n\}$ of order $k$ is defined to be the smallest positive number such that
\begin{align}
(1-\delta_{k|\mathcal{I}})\lVert x\rVert_2^p\leq\lVert Ax\rVert_p^p\leq (1+\delta_{k|\mathcal{I}})\lVert x\rVert_2^p
\end{align}
for all $x\in\mathbb{R}^N$ that are block $k$-sparse over $\mathcal{I}$.
\end{definition}

For convenience, we write $\delta_k$ for the block $p$-RIC $\delta_{k|\mathcal{I}}$ whenever there is no confusion. Then, we have the sufficient condition for the robust recovery result by using (\ref{min}) with the block $p$-RIC $\delta_k$.\\

\begin{theorem} Let $x\in\mathbb{R}^N$, and  $x^k$ be its best approximation with $k$ nonzero blocks, supported on block index set $T_0$. Let $\tilde{T}\subset\{1,2,\cdots,n\}$ be an arbitrary set. Define $\rho$ and $\alpha$ as before such that $|\tilde{T}|=\rho k$ and $|\tilde{T}\cap T_0|=\alpha\rho k$. Suppose that there exists an $a\in\mathbb{Z}$, with $a\geq (1-\alpha)\rho$, $a>1$, and the measurement matrix $A$ satisfies \begin{align}
\delta_{ak}+\frac{a^{1-p/2}}{\gamma}\delta_{(a+1)k}<\frac{a^{1-p/2}}{\gamma}-1, \label{bprip}
\end{align}
where $\gamma=\omega+(1-\omega)(1+\rho-2\alpha\rho)^{1-p/2}$ for some given $0\leq \omega\leq 1$. Then the solutions $x^{\sharp}$ of problem (\ref{min}) obeys
\begin{align}
\lVert x^{\sharp}-x\rVert_2\leq C_1\frac{\Big(\omega\lVert x-x^k\rVert_{2,p}^p+(1-\omega)\lVert x_{\tilde{T}^c\cap T_0^c}\rVert_{2,p}^p\Big)^{1/p}}{k^{1/p-1/2}}+C_2\varepsilon,
\end{align}
for some positive constants $C_1$ and $C_2$. 
\end{theorem}

\noindent
{\bf Remark 1.} Note that this theorem involves two important ratios: $\rho$ determines the ratio of the estimated block support size to the actual block support of $x^k$ (or the block support of $x$ if $x$ is block $k$-sparse), while $\alpha$ determines the ratio of the number of block indices in block support of $x^k$ that were accurately estimated in $\tilde{T}$ to the size of $\tilde{T}$. Specifically, $\alpha=\frac{|\tilde{T}\cap T_0|}{|\tilde{T}|}$.\\

\noindent
{\bf Remark 2.} The constants $C_1$ and $C_2$ are explicitly given by the following expressions:
\begin{align*}
C_1&=\frac{2^{2/p-1}a^{1/2-1/p}\left[(1+\delta_{ak})^{1/p}+(1-\delta_{(a+1)k})^{1/p}\right]}{[(1-\delta_{(a+1)k})-a^{p/2-1}(1+\delta_{ak})\gamma]^{1/p}},\\
C_2&=\frac{2^{1/p}m^{1/p-1/2}(1+a^{1/2-1/p}\gamma^{1/p})}{[(1-\delta_{(a+1)k})-a^{p/2-1}(1+\delta_{ak})\gamma]^{1/p}}.
\end{align*} \\

\noindent
{\bf Remark 3.} For Theorem 1 to be held, it is sufficient that $A$ satisfies \begin{align}
\delta_{(a+1)k}<\delta:=\frac{a^{1-p/2}-[\omega+(1-\omega)(1+\rho-2\alpha\rho)^{1-p/2}]}{a^{1-p/2}+[\omega+(1-\omega)(1+\rho-2\alpha\rho)^{1-p/2}]}. \label{delta}
\end{align}\\

Next, we illustrate how the slighted stronger sufficient conditions (\ref{delta}) vary with $\alpha$ and $\omega$ for $p=0.01,\,0.5,\,1$, respectively. In Figure 1, for each $p$, we plot $\delta$ versus $\omega$ for different values of $\alpha$, where we set the parameters $a=3$ and $\rho=1$. We observe that as $\alpha$ increases the sufficient condition on the block $p$-RIC becomes weaker ($\delta$ getting larger), allowing for a wider class of measurement matrices $A$. For each $p$ with $\alpha>0.5$, the sufficient condition becomes stronger as $\omega$ increases. For instance, when $\alpha=0.7$, $p=0.5$, with $\omega=0.2$ it suffices to have $\delta=0.5072$, compared with $\delta=0.3902$ for $\omega=1$. The opposite conclusion holds for the case $\alpha<0.5$. When $\alpha=0.5$, the sufficient condition remains the same for different $\omega$. From another point of view, as $p$ decreases, the sufficient condition becomes weaker, which reflects the benefits of using nonconvex minimization.  \\

\begin{figure}[htbp]
	\centering
	\includegraphics[width=\textwidth,height=0.4\textheight]{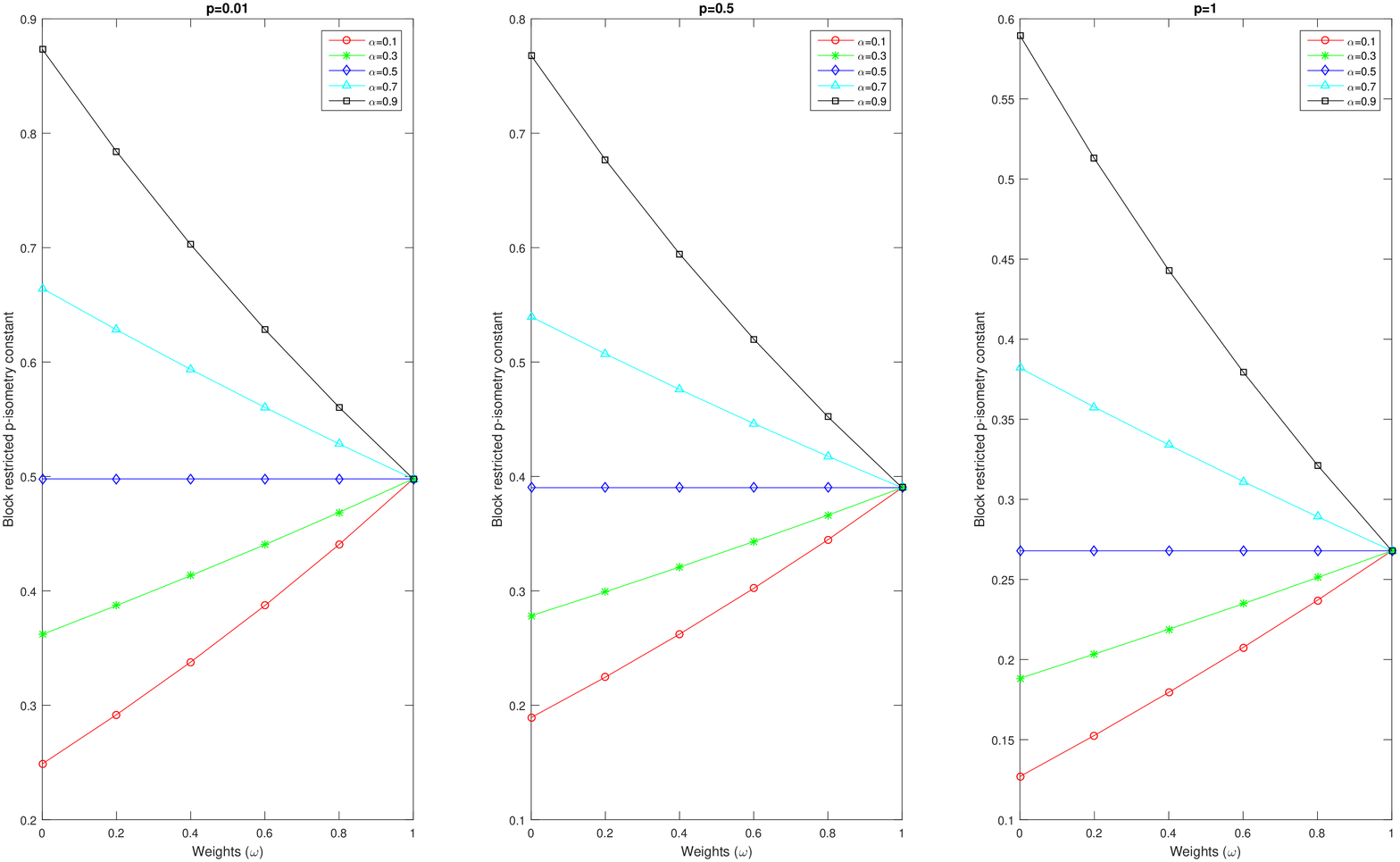}
	\caption{Comparison of the sufficient conditions (\ref{delta}) for the block restricted $p$-isometry constants with various of $\alpha$ for $p=0.01,\,0.5,\,1$. We set $a=3$ and $\rho=1$ for all the figures.}\label{fig:1}
\end{figure}

\noindent
{\bf Remark 4.} By setting $\omega=1$ and $a=b^{2/(2-p)}$ with $b>1$, this theorem reduces to Theorem 2.1 in \cite{wwx2}. In addition, by setting  $\omega=0$, $\alpha=0.5$ and $a=\rho=b^{2/(2-p)}$ with $b>1$, then this theorem goes to Theorem 2 in \cite{hwwx} with $s=ak$.\\

\noindent
{\bf Proof.}\,\, Let $x^{\sharp}=x+h$ be a solution of problem (\ref{min}), where $x$ is the unknown true signal and $h$ is the approximation error. Throughout the paper, $x_T$ will denote the vector equal to $x$ on the block index set $T$ and zero elsewhere. Then, we have \begin{align*}
\omega\lVert x^{\sharp}_{\tilde{T}}\rVert_{2,p}^p+\lVert x^{\sharp}_{\tilde{T}^c}\rVert_{2,p}^p\leq \omega\lVert x_{\tilde{T}}\rVert_{2,p}^p+\lVert x_{\tilde{T}^c}\rVert_{2,p}^p.
\end{align*}
That is, \begin{align*}
\omega\lVert x_{\tilde{T}}+h_{\tilde{T}}\rVert_{2,p}^p+\lVert x_{\tilde{T}^c}+h_{\tilde{T}^c}\rVert_{2,p}^p\leq \omega\lVert x_{\tilde{T}}\rVert_{2,p}^p+\lVert x_{\tilde{T}^c}\rVert_{2,p}^p.
\end{align*}
Consequently, \begin{align*}
&\omega\lVert x_{\tilde{T}\cap T_0}+h_{\tilde{T}\cap T_0}\rVert_{2,p}^p+\omega\lVert x_{\tilde{T}\cap T_0^c}+h_{\tilde{T}\cap T_0^c}\rVert_{2,p}^p+\lVert x_{\tilde{T}^c\cap T_0}+h_{\tilde{T}^c\cap T_0}\rVert_{2,p}^p+\lVert x_{\tilde{T}^c\cap T_0^c}+h_{\tilde{T}^c\cap T_0^c}\rVert_{2,p}^p \\
&\leq \omega\lVert x_{\tilde{T}\cap T_0}\rVert_{2,p}^p+\omega\lVert x_{\tilde{T}\cap T_0^c}\rVert_{2,p}^p+\lVert x_{\tilde{T}^c\cap T_0}\rVert_{2,p}^p+\lVert x_{\tilde{T}^c\cap T_0^c}\rVert_{2,p}^p.
\end{align*}
The forward and reverse triangle inequalities implies \begin{align*}
\omega\lVert h_{\tilde{T}\cap T_0^c}\rVert_{2,p}^p+\lVert h_{\tilde{T}^c\cap T_0^c}\rVert_{2,p}^p\leq \lVert h_{\tilde{T}^c\cap T_0}\rVert_{2,p}^p+\omega\lVert h_{\tilde{T}\cap T_0}\rVert_{2,p}^p+2\Big(\lVert x_{\tilde{T}^c\cap T_0^c}\rVert_{2,p}^p+\omega\lVert x_{\tilde{T}\cap T_0^c}\rVert_{2,p}^p\Big).
\end{align*}
Adding and subtracting $\omega\lVert h_{\tilde{T}^c\cap T_0^c}\rVert_{2,p}^p$ on the left hand side, and $\omega\lVert h_{\tilde{T}^c\cap T_0}\rVert_{2,p}^p$ on the right, we obtain \begin{align*}
\omega\lVert h_{T_0^c}\rVert_{2,p}^p+(1-\omega)\lVert h_{\tilde{T}^c\cap T_0^c}\rVert_{2,p}^p\leq \omega\lVert h_{T_0}\rVert_{2,p}^p+(1-\omega)\lVert h_{\tilde{T}^c\cap T_0}\rVert_{2,p}^p+2\Big(\omega\lVert x_{T_0^c}\rVert_{2,p}^p+(1-\omega)\lVert x_{\tilde{T}^c\cap T_0^c}\rVert_{2,p}^p\Big).
\end{align*}
But we can also write $$
\lVert h_{T_0^c}\rVert_{2,p}^p=\omega\lVert h_{T_0^c}\rVert_{2,p}^p+(1-\omega)\lVert h_{\tilde{T}^c\cap T_0^c}\rVert_{2,p}^p+(1-\omega)\lVert h_{\tilde{T}\cap T_0^c}\rVert_{2,p}^p.
$$
Therefore, we have $$
\lVert h_{T_0^c}\rVert_{2,p}^p\leq \omega\lVert h_{T_0}\rVert_{2,p}^p+(1-\omega)\lVert h_{\tilde{T}\cap T_0^c}\rVert_{2,p}^p+(1-\omega)\lVert h_{\tilde{T}^c\cap T_0}\rVert_{2,p}^p+2\Big(\omega\lVert x_{T_0^c}\rVert_{2,p}^p+(1-\omega)\lVert x_{\tilde{T}^c\cap T_0^c}\rVert_{2,p}^p\Big).
$$
Let the set $\tilde{T}_\alpha=T_0\cap\tilde{T}$, then we can write $\lVert h_{\tilde{T}\cap T_0^c}\rVert_{2,p}^p+\lVert h_{\tilde{T}^c\cap T_0}\rVert_{2,p}^p=\lVert h_{(T_0\cup\tilde{T})\setminus\tilde{T}_\alpha }\rVert_{2,p}^p$ and simplify the bound on \begin{align}
\lVert h_{T_0^c}\rVert_{2,p}^p\leq \omega\lVert h_{T_0}\rVert_{2,p}^p+(1-\omega)\lVert h_{(T_0\cup\tilde{T})\setminus\tilde{T}_\alpha }\rVert_{2,p}^p+2\Big(\omega\lVert x_{T_0^c}\rVert_{2,p}^p+(1-\omega)\lVert x_{\tilde{T}^c\cap T_0^c}\rVert_{2,p}^p\Big).\label{T0}
\end{align}

Next, we decompose $h_{T_0^c}$ into disjoint block index set $T_j$, each of $T_j\,(j\geq 1)$ consists of $ak$ blocks, where $a>1$. That is, $T_1$ indexes the $ak$ blocks with largest $\ell_2$ norm of $h_{T_0^c}$, $T_2$ indexes the second $ak$ blocks with largest $\ell_2$ norm of $h_{(T_0\cup T_1)^c}$, and so on. This gives $h_{T_0^c}=\sum\limits_{j\geq 1}h_{T_j}$. For each $i\in T_j\,(j\geq 2)$, it is easy to see that $$
\lVert h_{T_j}[i]\rVert_2^p\leq \frac{\lVert h_{T_{j-1}}[i]\rVert_2^p+\cdots+\lVert h_{T_{j-1}}[ak]\rVert_2^p}{ak}=\frac{\lVert h_{T_{j-1}}\rVert_{2,p}^p}{ak}.
$$
Then \begin{align*}
& \lVert h_{T_j}[i]\rVert_2^2\leq \frac{\lVert h_{T_{j-1}}\rVert_{2,p}^2}{(ak)^{2/p}},\,\,\, \lVert h_{T_j}\rVert_2^2\leq\frac{ak\lVert h_{T_{j-1}}\rVert_{2,p}^2}{(ak)^{2/p}}, \\
& \lVert h_{T_j}\rVert_2^p\leq\frac{\lVert h_{T_{j-1}}\rVert_{2,p}^p}{(ak)^{1-p/2}}.
\end{align*}
Thus, $$
\lVert h_{(T_0\cup T_1)^c}\rVert_2\leq \lVert \sum\limits_{j\geq 2}h_{T_j}\rVert_2\leq \sum\limits_{j\geq 2}\lVert h_{T_j}\rVert_2 \leq (ak)^{1/2-1/p}\sum\limits_{j\geq 2}\lVert h_{T_{j-1}}\rVert_{2,p}=(ak)^{1/2-1/p}\sum\limits_{j\geq 1}\lVert h_{T_{j}}\rVert_{2,p}.
$$
Hence, $$
\lVert h_{(T_0\cup T_1)^c}\rVert_2^p\leq (ak)^{p/2-1}\left(\sum\limits_{j\geq 1}\lVert h_{T_{j}}\rVert_{2,p}\right)^p\leq (ak)^{p/2-1}\sum\limits_{j\geq 1}\lVert h_{T_{j}}\rVert_{2,p}^p=(ak)^{p/2-1}\lVert h_{T_0^c}\rVert_{2,p}^p.
$$
Combining the above expression with (\ref{T0}), we get \begin{align}
\lVert h_{(T_0\cup T_1)^c}\rVert_2^p\leq (ak)^{p/2-1}\Bigg[ \omega\lVert h_{T_0}\rVert_{2,p}^p+(1-\omega)\lVert h_{(T_0\cup\tilde{T})\setminus\tilde{T}_\alpha }\rVert_{2,p}^p+2\Big(\omega\lVert x_{T_0^c}\rVert_{2,p}^p+(1-\omega)\lVert x_{\tilde{T}^c\cap T_0^c}\rVert_{2,p}^p\Big)\Bigg]. \label{T0T1C}
\end{align}

By H\"{o}lder's inequality it follows that 
$$
\lVert Ah\rVert_p^p\leq \left(\sum\limits_{i=1}^m(|(Ah)_i|^p)^{2/p}\right)^{p/2}\cdot\left(\sum\limits_{i=1}^m 1\right)^{1-p/2}=m^{1-p/2}\lVert Ah\rVert_2^p.
$$
Since $x^{\sharp}$ is a solution of problem (\ref{min}) and $\lVert y-Ax\rVert_2=\lVert e\rVert_2\leq \varepsilon$, so we have $$
\lVert Ah\rVert_2=\lVert A(x^{\sharp}-x)\rVert_2\leq \lVert Ax-y\rVert_2+\lVert Ax^{\sharp}-y\rVert_2\leq 2\varepsilon.
$$
Thus, $\lVert Ah\rVert_p^p\leq m^{1-p/2}\lVert Ah\rVert_2^p\leq m^{1-p/2}(2\varepsilon)^p$. Therefore, we obtain \begin{align*}
\lVert Ah_{T_0\cup T_1}\rVert_p^p&\leq \lVert Ah\rVert_p^p+\lVert Ah_{(T_0\cup T_1)^c}\rVert_p^p
\leq m^{1-p/2}(2\varepsilon)^p+\sum\limits_{j\geq 2}\lVert Ah_{T_j}\rVert_p^p\\
&\leq m^{1-p/2}(2\varepsilon)^p+(1+\delta_{ak})\sum\limits_{j\geq 2}\lVert h_{T_j}\rVert_2^p.
\end{align*}
Moreover, we have $$
\sum\limits_{j\geq 2}\lVert h_{T_j}\rVert_2^p\leq (ak)^{p/2-1}\sum\limits_{j\geq 2}\lVert h_{T_{j-1}}\rVert_{2,p}^p\leq (ak)^{p/2-1}\lVert h_{T_0^c}\rVert_{2,p}^p.
$$
Hence \begin{align*}
\lVert Ah_{T_0\cup T_1}\rVert_p^p&\leq m^{1-p/2}(2\varepsilon)^p+(ak)^{p/2-1}(1+\delta_{ak})\lVert h_{T_0^c}\rVert_{2,p}^p\\
&\leq m^{1-p/2}(2\varepsilon)^p+2(ak)^{p/2-1}(1+\delta_{ak})\Big(\omega\lVert x_{T_0^c}\rVert_{2,p}^p+(1-\omega)\lVert x_{\tilde{T}^c\cap T_0^c}\rVert_{2,p}^p\Big)\\
&\quad+\omega(ak)^{p/2-1}(1+\delta_{ak})\lVert h_{T_0}\rVert_{2,p}^p+(1-\omega)(ak)^{p/2-1}(1+\delta_{ak})\lVert h_{(T_0\cup\tilde{T})\setminus\tilde{T}_\alpha }\rVert_{2,p}^p.
\end{align*}
Noting that $|(T_0\cup\tilde{T})\setminus\tilde{T}_\alpha|=(1+\rho-2\alpha\rho)k$, we obtain $\lVert h_{(T_0\cup\tilde{T})\setminus\tilde{T}_\alpha }\rVert_{2,p}^p\leq \left[(1+\rho-2\alpha\rho)k\right]^{1-p/2}\lVert h_{(T_0\cup\tilde{T})\setminus\tilde{T}_\alpha }\rVert_{2}^p$.
Since the set $T_1$ contains the $ak$ blocks with largest $\ell_2$ norm of $h_{T_0^c}$ with $a>1$, and $|\tilde{T}\setminus \tilde{T}_\alpha|=(1-\alpha)\rho k\leq ak$, it follows that $\lVert h_{(T_0\cup\tilde{T})\setminus\tilde{T}_\alpha }\rVert_{2}^p\leq \lVert h_{T_0\cup T_1}\rVert_{2}^p$, which implies that $\lVert h_{(T_0\cup\tilde{T})\setminus\tilde{T}_\alpha }\rVert_{2,p}^p\leq\left[(1+\rho-2\alpha\rho)k\right]^{1-p/2}\lVert h_{T_0\cup T_1}\rVert_{2}^p$. In addition, $\lVert h_{T_0}\rVert_{2,p}^p\leq k^{1-p/2}\lVert h_{T_0}\rVert_{2}^p\leq k^{1-p/2}\lVert h_{T_0\cup T_1}\rVert_{2}^p$. Thus, \begin{align*}
(1-\delta_{(a+1)k})\lVert h_{T_0\cup T_1}\rVert_2^p&\leq \lVert Ah_{T_0\cup T_1}\rVert_p^p \\
&\leq  m^{1-p/2}(2\varepsilon)^p+2(ak)^{p/2-1}(1+\delta_{ak})\Big(\omega\lVert x_{T_0^c}\rVert_{2,p}^p+(1-\omega)\lVert x_{\tilde{T}^c\cap T_0^c}\rVert_{2,p}^p\Big) \\
&\quad+\omega a^{p/2-1}(1+\delta_{ak})\lVert h_{T_0\cup T_1}\rVert_{2}^p+(1-\omega)a^{p/2-1}(1+\rho-2\alpha\rho)^{1-p/2}(1+\delta_{ak})\lVert h_{T_0\cup T_1}\rVert_{2}^p.
\end{align*}
Therefore, if $(1-\delta_{(a+1)k})-a^{p/2-1}(1+\delta_{ak})\gamma>0$, that is $ \delta_{ak}+\frac{a^{1-p/2}}{\gamma}\delta_{(a+1)k}<\frac{a^{1-p/2}}{\gamma}-1$, then we have \begin{align}
\lVert h_{T_0\cup T_1}\rVert_2^p\leq \frac{m^{1-p/2}(2\varepsilon)^p+2(ak)^{p/2-1}(1+\delta_{ak})\Big(\omega\lVert x_{T_0^c}\rVert_{2,p}^p+(1-\omega)\lVert x_{\tilde{T}^c\cap T_0^c}\rVert_{2,p}^p\Big)}{(1-\delta_{(a+1)k})-a^{p/2-1}(1+\delta_{ak})\gamma},
\end{align}
where $\gamma=\omega+(1-\omega)(1+\rho-2\alpha\rho)^{1-p/2}$. According to (\ref{T0T1C}), we also have \begin{align*}
\lVert h_{(T_0\cup T_1)^c}\rVert_2^p&\leq a^{p/2-1}\gamma\lVert h_{T_0\cup T_1}\rVert_2^p+2(ak)^{p/2-1}\Big(\omega\lVert x_{T_0^c}\rVert_{2,p}^p+(1-\omega)\lVert x_{\tilde{T}^c\cap T_0^c}\rVert_{2,p}^p\Big)\\
&\leq \frac{a^{p/2-1}\gamma m^{1-p/2}(2\varepsilon)^p+2(ak)^{p/2-1}(1-\delta_{(a+1)k})\Big(\omega\lVert x_{T_0^c}\rVert_{2,p}^p+(1-\omega)\lVert x_{\tilde{T}^c\cap T_0^c}\rVert_{2,p}^p\Big)}{(1-\delta_{(a+1)k})-a^{p/2-1}(1+\delta_{ak})\gamma}.
\end{align*}
From $\lVert v\rVert_p\leq 2^{1/p-1}\lVert v\rVert_1$ for all $v\in \mathbb{R}^2$, it follows that  \begin{align*}
\lVert h\rVert_2&\leq \lVert h_{T_0\cup T_1}\rVert_2+ \lVert h_{(T_0\cup T_1)^c}\rVert_2 \\
&\leq 2^{1/p-1}\left(\frac{2m^{1/p-1/2}\varepsilon+2^{1/p}(ak)^{1/2-1/p}(1+\delta_{ak})^{1/p}\Big(\omega\lVert x_{T_0^c}\rVert_{2,p}^p+(1-\omega)\lVert x_{\tilde{T}^c\cap T_0^c}\rVert_{2,p}^p\Big)^{1/p}}{[(1-\delta_{(a+1)k})-a^{p/2-1}(1+\delta_{ak})\gamma]^{1/p}}\right)\\
&\quad+2^{1/p-1}\left(\frac{2a^{1/2-1/p}\gamma^{1/p}m^{1/p-1/2}\varepsilon+2^{1/p}(ak)^{1/2-1/p}(1-\delta_{(a+1)k})^{1/p}\Big(\omega\lVert x_{T_0^c}\rVert_{2,p}^p+(1-\omega)\lVert x_{\tilde{T}^c\cap T_0^c}\rVert_{2,p}^p\Big)^{1/p}}{[(1-\delta_{(a+1)k})-a^{p/2-1}(1+\delta_{ak})\gamma]^{1/p}}\right)\\
&\leq \frac{2^{2/p-1}(ak)^{1/2-1/p}\left[(1+\delta_{ak})^{1/p}+(1-\delta_{(a+1)k})^{1/p}\right]\Big(\omega\lVert x_{T_0^c}\rVert_{2,p}^p+(1-\omega)\lVert x_{\tilde{T}^c\cap T_0^c}\rVert_{2,p}^p\Big)^{1/p}}{[(1-\delta_{(a+1)k})-a^{p/2-1}(1+\delta_{ak})\gamma]^{1/p}} \\
&\quad+\frac{2^{1/p}m^{1/p-1/2}(1+a^{1/2-1/p}\gamma^{1/p})\varepsilon}{[(1-\delta_{(a+1)k})-a^{p/2-1}(1+\delta_{ak})\gamma]^{1/p}}=C_1\frac{\Big(\omega\lVert x_{T_0^c}\rVert_{2,p}^p+(1-\omega)\lVert x_{\tilde{T}^c\cap T_0^c}\rVert_{2,p}^p\Big)^{1/p}}{k^{1/p-1/2}}+C_2\varepsilon,
\end{align*}
which completes the proof.\\

\begin{cor} Let $A\in\mathbb{R}^{m\times N}$ be a measurement matrix, $x\in\mathbb{R}^N$ be a block $k$-sparse signal supported on block index set $T_0$ with $y=Ax$, and $0<p\leq 1$. Let $\tilde{T}\subset\{1,2,\cdots,n\}$ be an arbitrary set. Define $\rho$ and $\alpha$ as before such that $|\tilde{T}|=\rho k$ and $|\tilde{T}\cap T_0|=\alpha\rho k$. Suppose that there exists an $a\in\mathbb{Z}$, with $a\geq (1-\alpha)\rho$, $a>1$, and the measurement matrix $A$ satisfies \begin{align}
\delta_{ak}+\frac{a^{1-p/2}}{\gamma}\delta_{(a+1)k}<\frac{a^{1-p/2}}{\gamma}-1,
\end{align}
where $\gamma=\omega+(1-\omega)(1+\rho-2\alpha\rho)^{1-p/2}$ for some given $0\leq \omega\leq 1$. Then the unique solution of problem (\ref{min}) with $\varepsilon=0$ is exactly $x$.
\end{cor}

\noindent
{\bf Remark 5.} According to the arguments in \cite{hwwx,wwx2}, if we let $A$ be an $m\times N$ matrix whose entries are i.i.d Gaussian random variables with mean zero, then there exist $C_3(p)$ and $C_4(p)$ such that whenever $0<p\leq 1$ and \begin{align}
m\geq C_3(p)kd+pC_4(p)k\ln(n/k), \label{number}
\end{align} the block $p$-RIP (\ref{bprip}) will hold and the block sparse signal $x$ can be exactly recovered with high probability. For a given $p\in(0,1]$, $C_3(p)$ and $C_4(p)$ are finite constants, and the second term of (\ref{number}) has the dominant impact on the number of measurements in an asymptotic sense. When $p\rightarrow 0$, the condition reduces to $m\geq C_3(0)kd$. While, when $p=1$, the required number of measurements $m\geq C_3(1)kd+C_4(1)k\ln(n/k)$, which implies fewer measurements are required with smaller $p$ for exact recovery via weighted mixed $\ell_2/\ell_p$ minimization than the case that $p=1$. This is largely consistent with what we have mentioned in Remark 3, namely, as $p$ decreases, the sufficient condition for the exact recovery becomes weaker. Though we have only considered the case that $d_1=d_2=\cdots=d_n=d$, the results here can be adapted to the case in which $d_i$ are not equal via replacing $d$ by the maximal block length $\max\limits_{1\leq i\leq n}d_i$.

\section{Weighted Block $p$-NSP}

In this section, we focus on the noise free case, that is $e=0$, and the signal $x$ is exactly block $k$-sparse, supported on block index set $T_0$. We obtain the exact recovery condition for the problem (\ref{min}) with $\varepsilon=0$ by using weighted block $p$-null space property. For an index set $V\subset\{1,\cdots,n\}$, we define $$
\Gamma_s(V):=\{U\subset\{1,\cdots,n\}: |(V\cap U^c)\cup (V^c\cap U)|\leq s\}.
$$
Basically, for any index set $V$, the set $(V\cap U^c)\cup (V^c\cap U)$ indicates the estimate error of $U$ with respective to $V$. So $\Gamma_s(V)$ includes all the index set estimates of $V$ that own an error of size $s$ or less. 
\begin{definition} Let $T\subset\{1,\cdots,n\}$ with $|T|\leq k$ and $\tilde{T}\in \Gamma_s(T)$. Assume the block size equals $d$. A matrix $A\in\mathbb{R}^{m\times N}$ is said to have the weighted nonuniform block $p$-null space property with parameters $T$ and $\tilde{T}$, and constant $C$ if for any vector $h: Ah=0$, we have \begin{align}
\omega\lVert h_T\rVert_{2,p}^p+(1-\omega)\lVert h_S\rVert_{2,p}^p\leq C\lVert h_{T^c}\rVert_{2,p}^p,
\end{align}
where $S=(\tilde{T}\cap T^c)\cup(\tilde{T}^c\cap T)$. In this case, we say $A$ satisfies $\omega$-$d$-$p$-$\mathrm{NSP}(T,\tilde{T},C)$.
\end{definition}

Next, we define a weighted uniform block $p$-null space property that leads to a sufficient and necessary condition for the exact recovery of all block $k$-sparse signals from compressive measurements using weighted mixed $\ell_2/\ell_p$ minimization problem.\\

\begin{definition} Assume the block size equals $d$. A matrix $A\in\mathbb{R}^{m\times N}$ is said to have the weighted block $p$-null space property with parameters $k$ and $s$, and constant $C$ if for any vector $h: Ah=0$, and for every block index set $T\subset\{1,\cdots,n\}$ with $|T|\leq k$ and $S\subset\{1,\cdots,n\}$ with $|S|\leq s$, we have \begin{align}
\omega\lVert h_T\rVert_{2,p}^p+(1-\omega)\lVert h_S\rVert_{2,p}^p\leq C\lVert h_{T^c}\rVert_{2,p}^p.
\end{align}
In this case, we say $A$ satisfies $\omega$-$d$-$p$-$\mathrm{NSP}(k,s,C)$.
\end{definition}

\noindent
{\bf Remark 6.} Our notations $\omega$-$d$-$p$-$\mathrm{NSP}(T,\tilde{T},C)$ and $\omega$-$d$-$p$-$\mathrm{NSP}(k,s,C)$ are direct extensions of the recent notations $\omega$-$\mathrm{NSP}(T,\tilde{T},C)$ and  $\omega$-$\mathrm{NSP}(k,s,C)$ proposed by \cite{ms}, which corresponds to the special case that $p=1$ and the block size $d=1$. That's to say $\omega$-1-1-$\mathrm{NSP}(k,s,C)$ is equivalent to $\omega$-$\mathrm{NSP}(k,s,C)$. Consequently, 1-1-1-$\mathrm{NSP}(k,k,C)$ is the standard null space property of order $k$, i.e, $\mathrm{NSP}(k,C)$.\\

The following theorem presents the sufficient and necessary condition for weighted mixed $\ell_2/\ell_p$ minimization problem to recover all block $k$-sparse signals when the error in the block support estimate is of size $s$ or less.

\begin{theorem}
Assume the block size equals $d$. Given a matrix $A\in\mathbb{R}^{m\times N}$, every block $k$-sparse signal $x\in\mathbb{R}^N$, supported on block index set $T_0$, is the unique solution of the weighted mixed $\ell_2/\ell_p$, $0<p\leq 1$ norm minimization problem (\ref{min}) with $\varepsilon=0$ and $\tilde{T}=\Gamma_s(T_0)$, if and only if $A$ satisfies $\omega$-$d$-$p$-$\mathrm{NSP}(k,s,C)$ for some positive constant $C<1$. 
\end{theorem}

\noindent
{\bf Proof.} \,\,a) $"\Rightarrow"$. Assume if $A$ does not satisfy $\omega$-$d$-$p$-$\mathrm{NSP}(k,s,C)$ for any constant $C>1$, then there exists a vector $h: Ah=0$ and block index set $T$ with $|T|\leq k$ and block index set $S$ with $|S|\leq s$, such that $Ah_T=-Ah_{T^c}$ and $$
\omega\lVert h_T\rVert_{2,p}^p+(1-\omega)\lVert h_S\rVert_{2,p}^p\geq \lVert h_{T^c}\rVert_{2,p}^p.
$$
Define $\tilde{T}=(T^c\cap S)\cup(T\cap S^c)$ so that $S=(T\cap\tilde{T}^c)\cup(T^c\cap\tilde{T})$. Substituting for $S$ and splitting the set $T$, we obtain \begin{align*}
&\omega\left(\lVert h_{T\cap \tilde{T}^c}\rVert_{2,p}^p+\lVert h_{T\cap\tilde{T}}\rVert_{2,p}^p\right)+(1-\omega)\left(\lVert h_{T\cap \tilde{T}^c}\rVert_{2,p}^p+\lVert h_{T^c\cap\tilde{T}}\rVert_{2,p}^p\right)\\
&=\lVert h_{T\cap \tilde{T}^c}\rVert_{2,p}^p+\omega\lVert h_{T\cap \tilde{T}}\rVert_{2,p}^p+(1-\omega)\lVert h_{T^c\cap \tilde{T}}\rVert_{2,p}^p\geq
\lVert h_{T^c}\rVert_{2,p}^p.
\end{align*}
Then, we have $$
\omega\lVert h_{T\cap \tilde{T}}\rVert_{2,p}^p+\lVert h_{T\cap \tilde{T}^c}\rVert_{2,p}^p\geq \omega\lVert h_{T^c\cap \tilde{T}}\rVert_{2,p}^p+\lVert h_{T^c\cap \tilde{T}^c}\rVert_{2,p}^p.
$$
In other words, the weighted mixed $\ell_2/\ell_p$ norm of the vector $h_T$ equals or exceeds that of $-h_{T^c}$. So $h_T$ is not the unique minimizer, which is contradictory to our condition. Thus, the necessity is verified. \\

b) $"\Leftarrow"$. Let $x^{\sharp}$ be a minimizer of  weighted mixed $\ell_2/\ell_p, 0<p\leq 1$ norm minimization problem (\ref{min}) with $\varepsilon=0$ and $\tilde{T}=\Gamma_s(T_0)$ and define $h=x^{\sharp}-x$. Then, by the optimality of $x^{\sharp}$, we have $$
\omega\lVert x_{\tilde{T}}+h_{\tilde{T}}\rVert_{2,p}^p+\lVert x_{\tilde{T}^c}+h_{\tilde{T}^c}\rVert_{2,p}^p\leq \omega\lVert x_{\tilde{T}}\rVert_{2,p}^p+
\lVert x_{\tilde{T}^c}\rVert_{2,p}^p,
$$
which is equivalent to \begin{align*}
&\omega\lVert x_{\tilde{T}\cap T_0}+h_{\tilde{T}\cap T_0}\rVert_{2,p}^p+\omega\lVert x_{\tilde{T}\cap T_0^c}+h_{\tilde{T}\cap T_0^c}\rVert_{2,p}^p+\lVert x_{\tilde{T}^c\cap T_0}+h_{\tilde{T}^c\cap T_0}\rVert_{2,p}^p+\lVert x_{\tilde{T}^c\cap T_0^c}+h_{\tilde{T}^c\cap T_0^c}\rVert_{2,p}^p \\
&\leq \omega\lVert x_{\tilde{T}\cap T_0}\rVert_{2,p}^p+ \omega\lVert x_{\tilde{T}\cap T_0^c}\rVert_{2,p}^p+\lVert x_{\tilde{T}^c\cap T_0}\rVert_{2,p}^p+\lVert x_{\tilde{T}^c\cap T_0^c}\rVert_{2,p}^p.
\end{align*}
Since $x$ is strictly block $k$-sparse and supported on the block index set $T_0$, thus $x_{T_0^c}=0$. Then we have \begin{align*}
\omega\lVert h_{\tilde{T}\cap T_0^c}\rVert_{2,p}^p+\lVert h_{\tilde{T}^c\cap T_0^c}\rVert_{2,p}^p\leq \lVert h_{\tilde{T}^c\cap T_0}\rVert_{2,p}^p+\omega\lVert h_{\tilde{T}\cap T_0}\rVert_{2,p}^p.
\end{align*}
Adding and subtracting  $\omega\lVert h_{\tilde{T}^c\cap T_0^c}\rVert_{2,p}^p$ on the left hand side, and $\omega\lVert h_{\tilde{T}^c\cap T_0}\rVert_{2,p}^p$ on the right, we obtain \begin{align*}
&\omega\lVert h_{\tilde{T}\cap T_0^c}\rVert_{2,p}^p+\omega\lVert h_{\tilde{T}^c\cap T_0^c}\rVert_{2,p}^p+\lVert h_{\tilde{T}^c\cap T_0^c}\rVert_{2,p}^p-\omega\lVert h_{\tilde{T}^c\cap T_0^c}\rVert_{2,p}^p \\
&\leq \omega\lVert h_{\tilde{T}\cap T_0}\rVert_{2,p}^p+\omega\lVert h_{\tilde{T}^c\cap T_0}\rVert_{2,p}^p+\lVert h_{\tilde{T}^c\cap T_0}\rVert_{2,p}^p-\omega\lVert h_{\tilde{T}^c\cap T_0}\rVert_{2,p}^p.
\end{align*}
Therefore, $$
\omega\lVert h_{T_0^c}\rVert_{2,p}^p+(1-\omega)\lVert h_{\tilde{T}^c\cap T_0^c}\rVert_{2,p}^p\leq \omega\lVert h_{T_0}\rVert_{2,p}^p+(1-\omega)\lVert h_{\tilde{T}^c\cap T_0}\rVert_{2,p}^p.
$$
Finally, by adding $(1-\omega)\lVert h_{\tilde{T}\cap T_0^c}\rVert_{2,p}^p$ to both sides, we have \begin{align}
\lVert h_{T_0^c}\rVert_{2,p}^p&\leq \omega\lVert h_{T_0}\rVert_{2,p}^p+(1-\omega)\lVert h_{\tilde{T}\cap T_0^c}\rVert_{2,p}^p+(1-\omega)\lVert h_{\tilde{T}^c\cap T_0^c}\rVert_{2,p}^p  \nonumber \\
&=\omega\lVert h_{T_0}\rVert_{2,p}^p+(1-\omega)\lVert h_S\rVert_{2,p}^p,
\end{align}
by setting $S=(\tilde{T}\cap T_0^c)\cup(\tilde{T}^c\cap T_0)$. Note that when $|S|\leq s$, the above inequality is in contradiction with that $A$ satisfies $\omega$-$d$-$p$-$\mathrm{NSP}(k,s,C)$ for some constant $C<1$, unless $h=0$. Hence, we have $x^{\sharp}=x$, and the sufficiency holds.

\section{From Block $p$-RIP to Weighted Block $p$-NSP}

In this part, we establish the relationship between the block $p$-RIP and weighted block $p$-NSP, which tells us that block $p$-RIP can directly imply the weighted block $p$-NSP. As a consequence, we can obtain the exact recovery for every block $k$-sparse signal via the weighted mixed $\ell_2/\ell_p, 0<p\leq 1$ norm minimization problem (\ref{min}) with $\varepsilon=0$. The relationship is obtained as follows.

\begin{theorem}  Assume the block size equals $d$ and let $x\in\mathbb{R}^N$ be block $k$-sparse, supported on block index set $T_0$. Let $\tilde{T}\subset\{1,2,\cdots,n\}$ be an arbitrary set and define $\rho$ and $\alpha$ as before such that $|\tilde{T}|=\rho k$ and $|\tilde{T}\cap T_0|=\alpha \rho k$. Let $s=(1+\rho-2\alpha\rho)k$, then we have $\tilde{T}=\Gamma_s(T_0)$. Suppose that there exists an $a\in\mathbb{Z}$, with $a\geq (1-\alpha)\rho$, $a>1$, and the measurement matrix $A$ satisfies \begin{align}
\delta_{ak}+\frac{a^{1-p/2}}{\gamma}\delta_{(a+1)k}<\frac{a^{1-p/2}}{\gamma}-1,
\end{align}
where $\gamma=\omega+(1-\omega)(s/k)^{1-p/2}$ for some given $0\leq \omega\leq 1$. Then $A$ satisfies $\omega$-$d$-$p$-$\mathrm{NSP}(k,s,C)$ for some constant $C<1$. 
\end{theorem}

Consequently, according to Theorem 2, we have $x$ is the unique solution of problem (\ref{min}) with $\varepsilon=0$.\\

\noindent
{\bf Proof.} To show for any vector $h: Ah=0$, and for every block index set $T\subset\{1,\cdots,n\}$ with $|T|\leq k$, and $S\subset\{1,\cdots,n\}$ with $|S|\leq s$, we have \begin{align}
\omega\lVert h_T\rVert_{2,p}^p+(1-\omega)\lVert h_S\rVert_{2,p}^p\leq C\lVert h_{T^c}\rVert_{2,p}^p, \label{nsp}
\end{align}
for some constant $C<1$, we only need to show (\ref{nsp}) holds for $T=G_0$ and $S=H_0$, where $G_0$ is the block index set over the $k$ blocks with largest $\ell_2$ norm of $h$, $H_0$ is the block index set over the $s$ blocks with largest $\ell_2$ norm of $h$.

Next, we decompose $h_{G_0^c}$ into disjoint block index set $G_j$, each of $G_{j}\,(j\geq 1)$ consists of $ak$ blocks. That is, $G_1$ indexes the $ak$ blocks with largest $\ell_2$ norm of $h_{G_0^c}$. $G_2$ indexes the second $ak$ blocks with largest $\ell_2$ norm of $h_{(G_0\cup G_1)^c}$, and so on.

Since $Ah=0$, then we have $$
\lVert A h_{G_0\cup G_1}\rVert_p^p\leq \lVert A h_{(G_0\cup G_1)^c}\rVert_p^p\leq \sum\limits_{j\geq 2}\lVert Ah_{G_j}\rVert_p^p\leq (1+\delta_{ak})\sum\limits_{j\geq 2}\lVert h_{G_j}\rVert_{2}^p.
$$
Moreover, we have $$
\sum\limits_{j\geq 2}\lVert h_{G_j}\rVert_{2}^p\leq (ak)^{p/2-1}\sum\limits_{j\geq 2}\lVert h_{G_{j-1}}\rVert_{2,p}^p\leq (ak)^{p/2-1}\lVert h_{G_0^c}\rVert_{2,p}^p.
$$
Thus, $$
(1-\delta_{(a+1)k})\lVert h_{G_0\cup G_1}\rVert_2^p\leq \lVert A h_{G_0\cup G_1}\rVert_p^p\leq (1+\delta_{ak})(ak)^{p/2-1}\lVert h_{G_0^c}\rVert_{2,p}^p.
$$
It implies that \begin{align}
\lVert h_{G_0\cup G_1}\rVert_2^p\leq \frac{(ak)^{p/2-1}(1+\delta_{ak})}{1-\delta_{(a+1)k}}\lVert h_{G_0^c}\rVert_{2,p}^p.
\end{align}
In addition, $\lVert h_{G_0}\rVert_{2,p}^p\leq k^{1-p/2}\lVert h_{G_0}\rVert_{2}^p\leq k^{1-p/2}\lVert h_{G_0\cup G_1}\rVert_{2}^p$. Hence, \begin{align}
\lVert h_{G_0}\rVert_{2,p}^p\leq k^{1-p/2}\lVert h_{G_0\cup G_1}\rVert_{2}^p\leq \frac{a^{p/2-1}(1+\delta_{ak})}{1-\delta_{(a+1)k}}\lVert h_{G_0^c}\rVert_{2,p}^p. \label{G0}
\end{align}

For the term $\lVert h_{H_0}\rVert_{2,p}^p$, we have $\lVert h_{H_0}\rVert_{2,p}^p\leq s^{1-p/2}\lVert h_{H_0}\rVert_{2}^p$. Moreover, if $s\leq k$, we have $H_0\subset G_0$ and $\lVert h_{H_0}\rVert_{2}^p\leq \lVert h_{G_0}\rVert_{2}^p\leq \lVert h_{G_0\cup G_1}\rVert_{2}^p $. If $s>k$, then $G_0\subset H_0$. However, since $|H_0\setminus G_0|=s-k=(1+\rho-2\alpha\rho)k-k=(1-2\alpha)\rho k\leq ak$, then we also have  $\lVert h_{H_0}\rVert_{2}^p\leq \lVert h_{G_0\cup G_1}\rVert_{2}^p$ as the block index set $G_1$ contains the $ak$ blocks with largest $\ell_2$ norm of $h_{G_0^c}$. As a consequence, \begin{align}
\lVert h_{H_0}\rVert_{2,p}^p\leq s^{1-p/2}\lVert h_{H_0}\rVert_{2}^p\leq s^{1-p/2}\lVert h_{G_0\cup G_1}\rVert_{2}^p\leq \frac{(ak/s)^{p/2-1}(1+\delta_{ak})}{1-\delta_{(a+1)k}}\lVert h_{G_0^c}\rVert_{2,p}^p.
\end{align}
Therefore, we obtain \begin{align*}
\omega\lVert h_{G_0}\rVert_{2,p}^p+(1-\omega)\lVert h_{H_0}\rVert_{2,p}^p&\leq \omega\frac{a^{p/2-1}(1+\delta_{ak})}{1-\delta_{(a+1)k}}\lVert h_{G_0^c}\rVert_{2,p}^p+(1-\omega)\frac{(ak/s)^{p/2-1}(1+\delta_{ak})}{1-\delta_{(a+1)k}}\lVert h_{G_0^c}\rVert_{2,p}^p \\
&=\frac{a^{p/2-1}\gamma(1+\delta_{ak})}{1-\delta_{(a+1)k}}\lVert h_{G_0^c}\rVert_{2,p}^p.
\end{align*}
Let $C=\frac{a^{p/2-1}\gamma(1+\delta_{ak})}{1-\delta_{(a+1)k}}$, we have $C<1$ under the condition of block $p$-RIP. The proof is completed.

\section{Simulation}
In this section, we conduct several simulations to illustrate our presented theoretical results. We adopt the iteratively reweighted least squares (IRLS) approach to solve the nonconvex optimization problem. We begin with $x^{(0)}=\arg\min\,\,\lVert y-Ax\rVert_2^2$, and set $\gamma_0=1$. Then, let $x^{(t+1)}$ be the solution of \begin{align}
\min\limits_{x}\frac{1}{2\lambda}\lVert y-Ax\rVert_2^2+\frac{1}{2}\lVert W^{(t)}x\rVert_2^2,
\end{align}
where $\lambda>0$ is a regularization parameter, and the weight matrix $W^{(t)}$ is defined as $W_i^{(t)}=\mathrm{diag}((\gamma_t+\omega_i^{\frac{4}{p(p-2)}}\lVert \omega_i^{1/p}x^{(t)}[i]\rVert_2^2)^{p/4-1/2})$ for the $i$-th block. Then, $x^{(t+1)}$ can be given explicitly as $$
x^{(t+1)}=(W^{(t)})^{-1}\Big([A(W^{(t)})^{-1}]^T[A(W^{(t)})^{-1}]+\lambda I\Big)^{-1}[A(W^{(t)})^{-1}]^T y.
$$
The value of $\gamma$ is decreased according to the rule $\gamma_{t+1}=0.1\gamma_{t}$, and the iteration is continued until $\lVert x^{(t+1)}-x^{(t)}\rVert_2\leq 10^{-5}$ or the iteration times is larger than $2500$. In our experiments, the measurement matrix $A$ is generated as an $m\times N$ matrix with entries drawing from i.i.d standard normal distribution. For a generated block sparse or nearly block sparse signal $x$, the measurements $y=Ax+\sigma z$ with $z$ being the standard Gaussian white noise. We consider several different parameter values to demonstrate the theoretical results. In each experiment, we report the average results over $20$ replications. For IRLS, we set $\lambda=10^{-6}$ in the noise free case ($\sigma=0$), and $\lambda=10^{-2}$ in the noisy case ($\sigma=0.01$). The recovery performance is assessed by the signal to noise ratio (SNR), which is given by $$
\mathrm{SNR}=20\log_{10}\left(\frac{\lVert x\rVert_2}{\lVert x-x^{\sharp}\rVert_2}\right).
$$

\subsection{Exactly block sparse case}
We first consider the case that $x$ is exactly block $k$-sparse with $k=20$. In this set of experiments, the signal is of length $N=400$, which is generated by choosing $k$ blocks uniformly at random, and then choosing the nonzero values from the standard normal distribution for these $k$ blocks.

Figure 2 illustrates the recovery performance of the weighted $\ell_2/\ell_p$ minimization problem with $p=0.5$ and $d=2$ in both with and without noise cases. It can be observed that when $\alpha=0.7$, the best recovery is achieved for very small $\omega$ whereas a $\omega=1$ results in the lowest SNR. On the other hand, when $\alpha\leq 0.5$, the performance of the recovery algorithms is better for large $\omega$ than that for small $\omega$. The case $\omega=0$ results in the lowest SNR. And as is expected that in all settings, comparing to the noise free case, we have a lower SNR in the noisy case. 

Figure 3 shows the averaged SNR using weighted mixed $\ell_2/\ell_p$ minimization for different values of the ratio parameter $\rho$ with fixed $p=0.5$ and $d=2$. It shows that when $\alpha=0.7$, using a larger support estimate results in better reconstruction. However, in both the noise free and noisy measurements cases, the recovery performance is more sensitive to the accuracy $\alpha$ of the block support estimate than the ratio parameter $\rho$.

In addition, we illustrate the impacts of $p$ and $d$ in Figure 4 for both the noise free and noisy measurements cases. We fix $k=20$, $\alpha=0.7$ and $\omega=0.5$. It is evident that as $p$ decreases, the recovery performance becomes better. And as $d$ increases, more number of measurements are required to obtain good reconstructions. These are largely consistent with the theoretical results in Section 2.

\begin{figure}[htp]
	\centering
	\includegraphics[width=\textwidth,height=0.4\textheight]{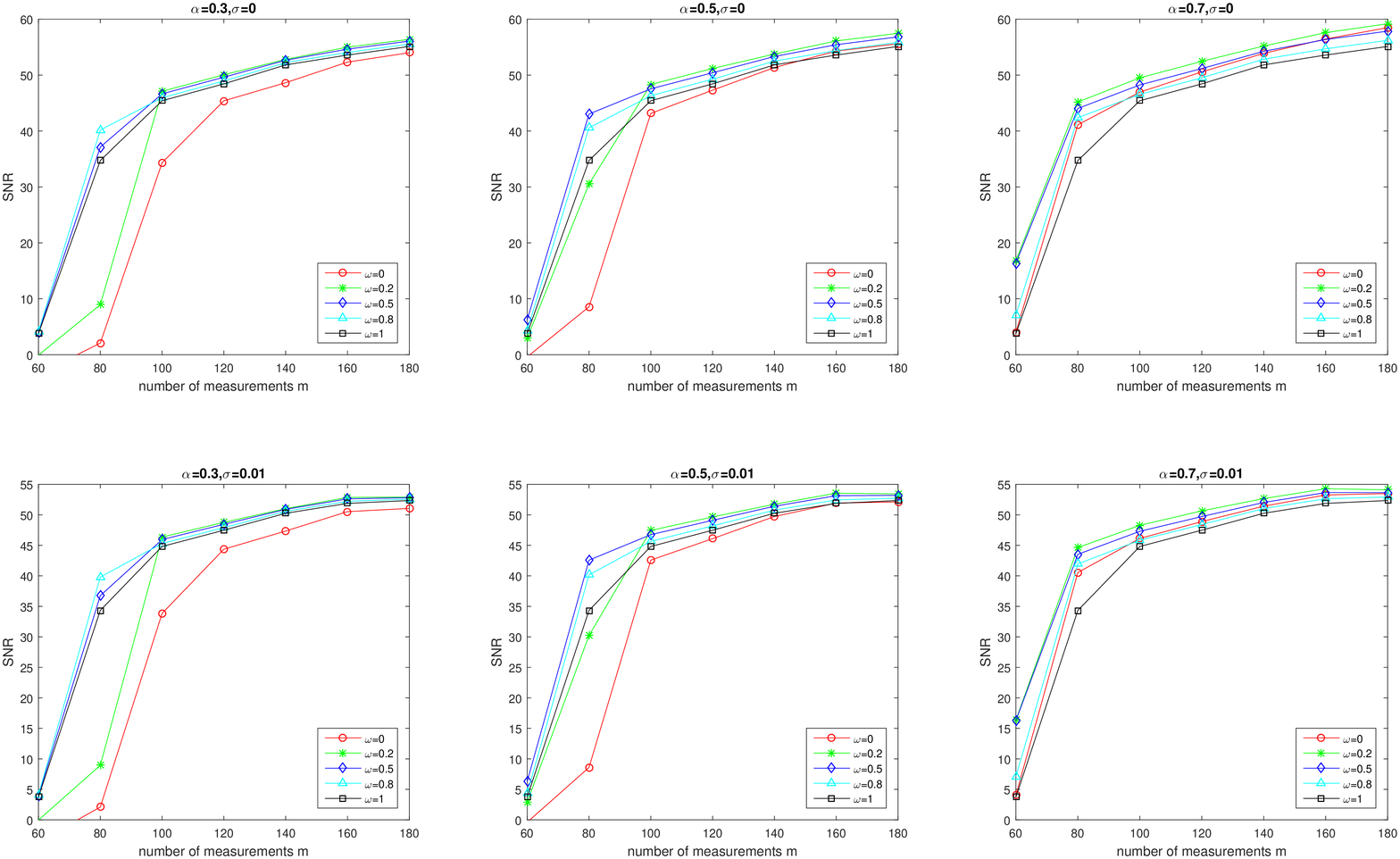}
	\caption{Performance of weighted mixed $\ell_2/\ell_p$ recovery with $p=0.5$ in terms of SNR for exactly block sparse signal $x$ depending on $\omega$ with $k=20$, $d=2$, while varying the number of measurements $m$.}\label{fig:2}
\end{figure}

\begin{figure}[htp]
	\centering
	\includegraphics[width=\textwidth,height=0.4\textheight]{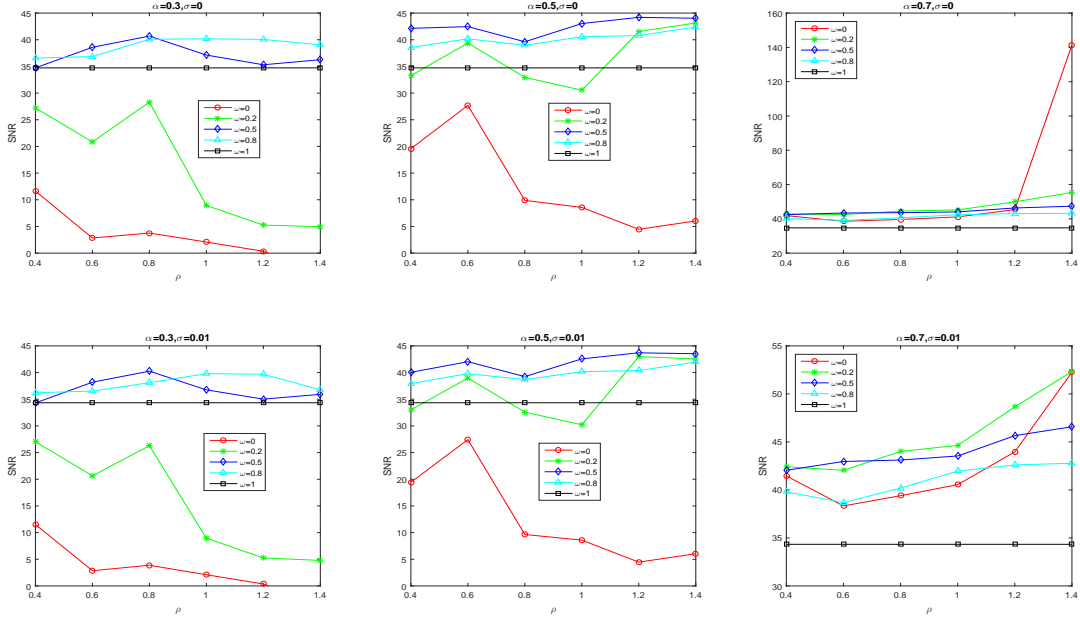}
	\caption{ Performance of weighted mixed $\ell_2/\ell_p$ recovery with $p=0.5$ in terms of SNR for exactly block sparse signal $x$ depending on $\omega$ with $k=20$, $d=2$, while varying the size of the block support estimate $\rho$ as a proportion of $k$.}\label{fig:3}
\end{figure}

\begin{figure}[htp]
	\centering
	\includegraphics[width=\textwidth,height=0.4\textheight]{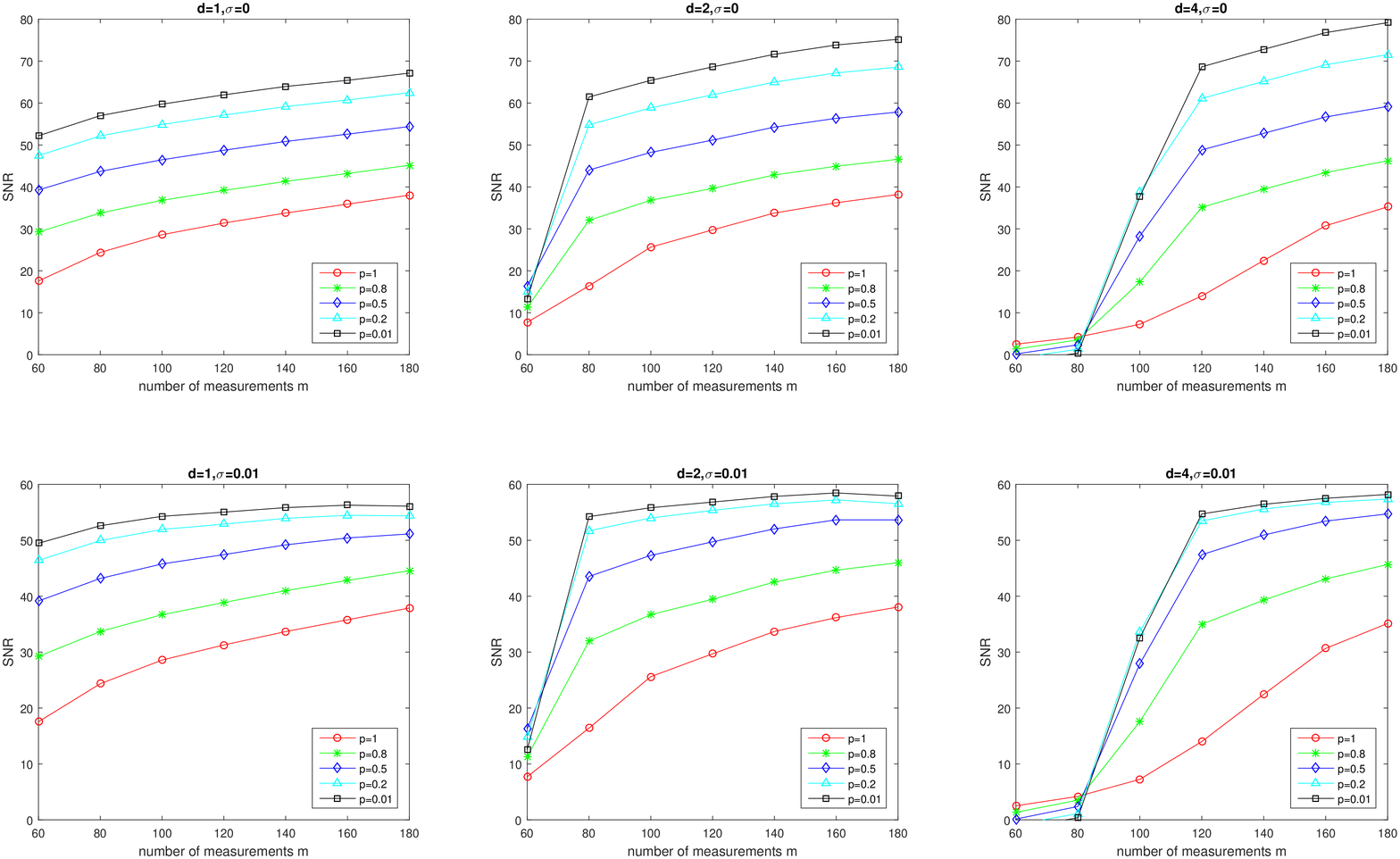}
	\caption{Performance of weighted mixed $\ell_2/\ell_p$ recovery in terms of SNR for exactly block sparse signal $x$ depending on $p$ with $k=20$, $\alpha=0.7$, $\omega=0.5$, while varying the number of measurements $m$.}\label{fig:4}
\end{figure}

\subsection{Nearly block sparse case}

Next, we generate $x$ with the $\ell_2$ norm of its blocks decay like $i^{-\theta}$ where $i\in\{1,\cdots,n\}$ and $\theta>1$. In Figure 5,  we illustrate the averaged SNR versus the size of the block support estimate $\rho$ as a proportion of $k$ for $\theta=1.5$. To calculate $\alpha$, we set $k=20$, i.e., we are interested in the best 20-term block sparse approximation. It shows that mediate values of $\omega$ (e.g., $\omega=0.2$ or 0.5) results in the best recovery. Generally, larger block support estimate favours better reconstruction result. Finally, we illustrate the impacts of $\theta$ and $d$ for different $p$ in Figure 6. For $\theta=1.5$, we set $k=20$, while for $\theta=2$, we consider $k=10$. It is evident that the recovery performance improves as $\theta$ increases, i.e., the block sparsity of signal increases. Moreover, decreasing $p$ and $d$ improves the recovery performance when other parameters are fixed.
\begin{figure}[htp]
	\centering
	\includegraphics[width=\textwidth,height=0.4\textheight]{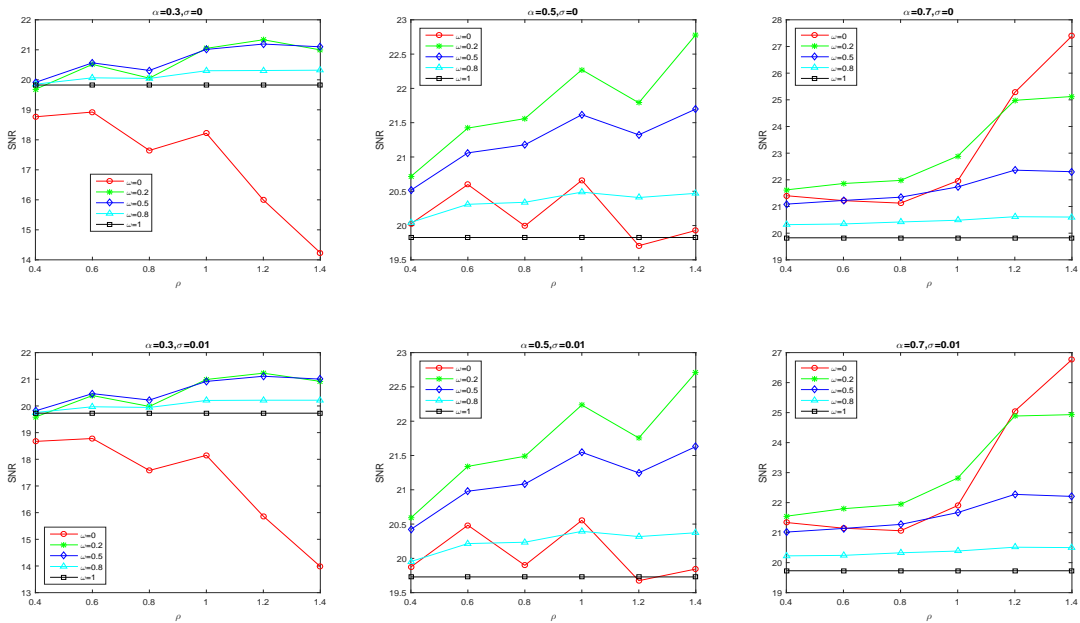}
	\caption{Performance of weighted mixed $\ell_2/\ell_p$ recovery in terms of SNR for nearly block sparse signal $x$ with $p=0.5$, $m=80$, $\theta=1.5$, $k=20$, $d=2$, while varying the size of the block support estimate $\rho$ as a proportion of $k$.}\label{fig:5}
\end{figure}

\begin{figure}[htp]
	\centering
	\includegraphics[width=\textwidth,height=0.4\textheight]{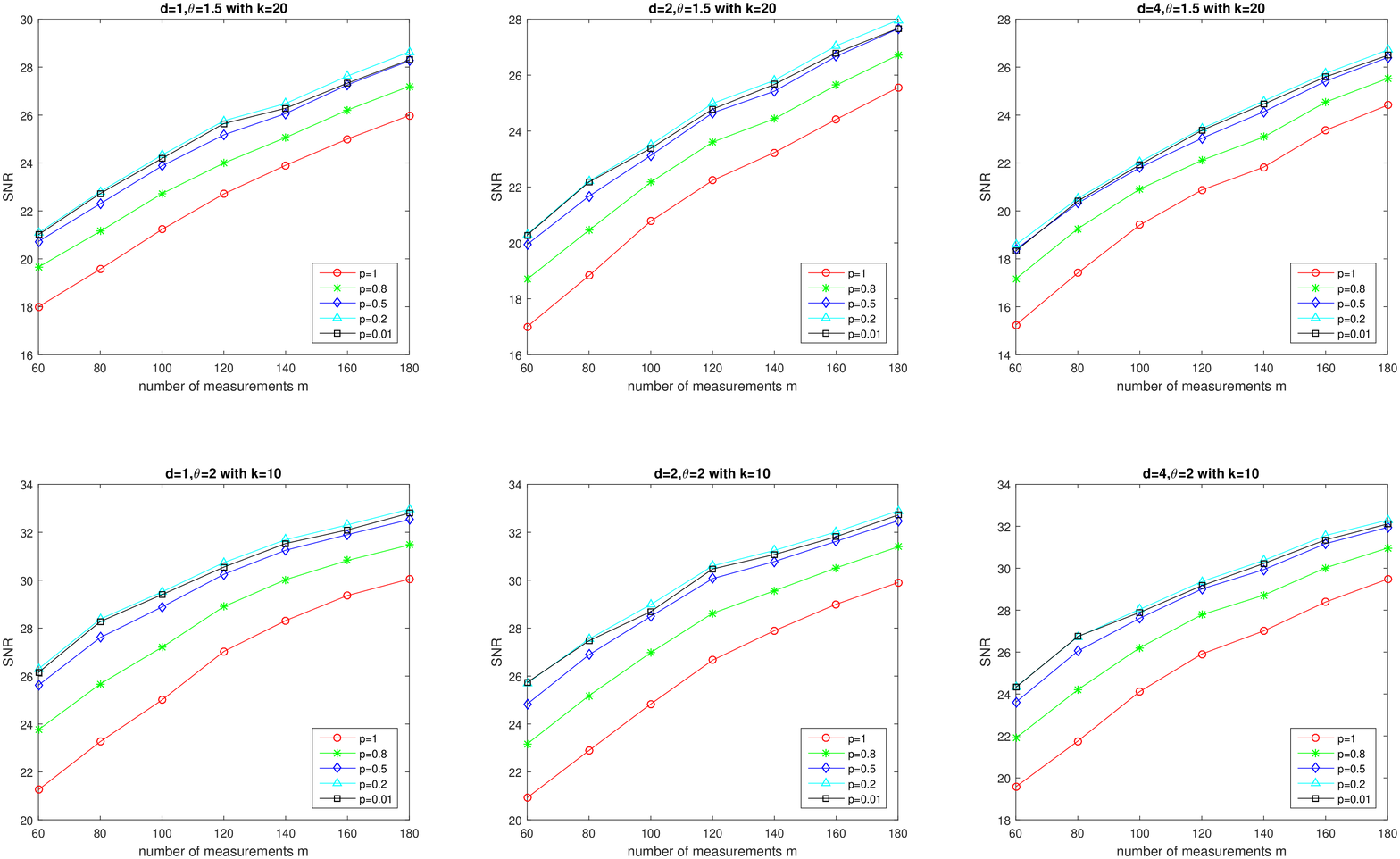}
	\caption{Performance of weighted mixed $\ell_2/\ell_p$ recovery in terms of SNR for nearly block sparse signal $x$ for different $\theta$ and $d$ depending on $p$ with fixed $\alpha=0.7$, $\omega=0.5$, $\sigma=0.01$ for all the cases, while varying the number of measurements $m$.}\label{fig:6}
\end{figure}

\section{Conclusion}
In this paper, we presented the recovery analysis of weighted mixed $\ell_2/\ell_p\,(0<p\leq 1)$ minimization by using both block $p$-RIP and weighted block $p$-null space property. In addition, we established the relationship between these two conditions. A series of simulations were conducted to illustrate our theoretical results. We largely generalized the existing results \cite{fmsy,hwwx,vl,wwx1,wwx2} to incorporating the prior known block support information with a weight $\omega\in[0,1]$. And from the simulation results, it shows that mediate values of the weight $\omega$ result in the best recovery in most of the cases. 

There are some interesting issues left for future work. One is that it might be possible to generalize $\ell_2$-constrained minimization problem considered in this paper to the $\ell_q$-constrained minimization problem with $2\leq q\leq \infty$ (see \cite{dlr,sac}) and with $0\leq q<2$ (see \cite{wllqy}). Another issue that is not addressed here is to obtain the minimum required number of measurements for perfect recovery directly via weighted block $p$-null space property. And the extension to the framework with completely arbitrary weights (see \cite{nsw}) needs also to be considered.

\section*{Acknowledgements}
 This work is supported by the Swedish Research Council grant (Reg.No. 340-2013-5342).

\end{document}